# TMEM240 protein, mutated in SCA21, is localized in Purkinje cells and synaptic terminals

Mégane Homa, MSc[a], Anne Loyens, BSc[a], Sabiha Eddarkaoui, MSc[a], Emilie Faivre, PhD[a], Vincent Deramecourt, MD, PhD[a], Claude-Alain Maurage, MD, PhD[c], Luc Buée, PhD[a], Vincent Huin, MD, PhD[a,b], Bernard Sablonnière, MD, PhD[a,b]

[a] Univ. Lille, Inserm, CHU Lille, UMR-S 1172 - JPArc - Centre de Recherche Jean-Pierre AUBERT Neurosciences et Cancer, F-59000, Lille, France

[b] CHU Lille, Institut de Biochimie et Biologie moléculaire, Centre de Biologie Pathologie et Génétique, F-59000, Lille, France

[c] CHU Lille, Laboratoire d'Anatomopathologie, Centre de Biologie Pathologie et Génétique, F-59000, Lille, France

**Correspondence to:**

Vincent HUIN: vincent.huin@inserm.fr

Inserm UMR-S 1172, JPArc, rue Polonovski, F-59045, Lille, France.

Telephone: +33-359-899-605

Fax: +33-320-538-562

**Running title:** TMEM240 localization in the cerebellum

**ORCID ID:**

Mégane Homa = https://orcid.org/0000-0002-3904-6888

Emilie Faivre = https://orcid.org/0000-0002-8608-9848

Vincent Deramecourt = https://orcid.org/0000-0001-9742-7833

Luc Buée = https://orcid.org/0000-0002-6261-4230

Vincent Huin = https://orcid.org/0000-0001-8201-5406

Claude-Alain Maurage = https://orcid.org/0000-0002-0229-6461

Bernard Sablonnière = https://orcid.org/0000-0003-0384-4076

## Abstract

A variety of missense mutations and a stop mutation in the gene coding for transmembrane protein 240 (*TMEM240*) have been reported to be the causative mutations of spinocerebellar ataxia 21 (SCA21). We aimed to investigate the expression of TMEM240 protein in mouse brain at the tissue, cellular, and subcellular levels. Immunofluorescence labeling showed TMEM240 to be expressed in various areas of the brain, with the highest levels in the hippocampus, isocortex, and cerebellum. In the cerebellum, TMEM240 was detected in the deep nuclei and the cerebellar cortex. The protein was expressed in all three layers of the cortex and various cerebellar neurons. TMEM240 was localized to climbing, mossy, and parallel fiber afferents projecting to Purkinje cells, as shown by coimmunostaining with VGLUT1 and VGLUT2. Co-immunostaining with synaptophysin, post-synaptic fractionation, and confirmatory electron microscopy showed TMEM240 to be localized to the post-synaptic side of synapses near the Purkinje-cell soma. Similar results were obtained in human cerebellar sections. These data suggest that TMEM240 may be involved in the organization of the cerebellar network, particularly in synaptic inputs converging on Purkinje cells. This study is the first to describe TMEM240 expression in the normal mouse brain.

# Introduction

Autosomal dominant cerebellar ataxias (SCAs) constitute a clinically and genetically heterogeneous group of diseases characterized by degeneration of the cerebellum, brainstem, and spinal cord [1]. Spinocerebellar ataxia type 21 (SCA21) is defined by mild but slowly progressive cerebellar ataxia that is frequently associated with varying degrees of cognitive impairment and Parkinsonism. We recently reported that SCA21 is caused by mutations in the transmembrane protein 240 (*TMEM240*) gene, previously called *C1orf70* [2, 3]. Six missense mutations and one stop mutation in the *TMEM240* gene have been described in patients in France, China, Japan, and, very recently, Germany, Colombia, and the Netherlands [2, 3, 4, 5, 6]. The *TMEM240* gene encodes a small transmembrane protein of unknown function, which has not yet been characterized. To date, non-targeted analyses in the mouse have detected TMEM240 protein in synaptic membranes of the brain [7]. *TMEM240* is transcribed in the adult human brain—particularly in the cerebral cortex, cerebellum, dentate gyrus, putamen, and caudate nucleus [8, 9]. Immunohistochemical staining of the human cerebellum has shown moderate positivity in Purkinje-cell membranes [9]. Although the cerebellum is well known to play a crucial role in movement coordination, balance, and vestibular control, a growing body of evidence has highlighted the structure's involvement in non-motor functions, such as cognition, language, and emotion [10, 11]. Cerebellar damage (including the damage observed in SCA) can be associated with cognitive impairment, such as that seen in cerebellar cognitive affective syndrome (CCAS) [12, 13]. In cases of SCA21, brain MRI has revealed atrophy of the vermis and the cerebellar hemispheres.

The anatomic structure of the cerebellum has been highly conserved throughout evolution. In mammals, the cerebellum is divided into 10 lobules separated by fissures of various depths [14]. The cerebellar cortex has a well-defined architecture comprising three different layers, each of which is composed of distinct but interconnected cell types. The external molecular layer contains not only inhibitory basket and stellate neurons but also climbing and parallel excitatory fibers that connect to Purkinje-cell dendrites. The Purkinje-cell layer consists of a single sheet of Purkinje-cell bodies, ascending Golgi-cell dendrites, descending basket cell dendrites, climbing fibers, and basket cell bodies. The internal granular layer contains a large number of granule cells. A significant proportion of the granular layer is occupied by glomeruli, which consist of dendrodendritic and axodendritic synapses mainly composed of mossy fibers and inhibitory Golgi-cell dendrites that converge on granule-cell dendrites [15]. The granular layer also contains Purkinje-cell axons leading to a broad network of fiber tracts that connect to deep nuclei (the fastigial, interposed, and dentate nuclei). These nuclei receive inhibitory inputs from Purkinje cells, as well as excitatory inputs from mossy and climbing fibers. The deep nuclei project numerous efferents to other parts of the brain and spinal cord [16]. Purkinje cells are the main targets in the cerebellar network; they receive and integrate excitatory and inhibitory inputs and transfer inhibitory signals to deep nuclei. There are two sources of afferents in the cerebellum: climbing and mossy fibers. Climbing fibers arise from the inferior olivary nucleus and pass through the granular layer to connect to Purkinje-cell bodies and dendrites. Mossy fibers arise in various areas of the brain and spinal cord, cross the white matter to reach the glomeruli, and target granule-cell dendrites. The axons of granule cells reach the molecular layer, where they form T-shaped branches, resulting in two branches of parallel fibers [17]. Here, we investigated TMEM240 protein expression in the adult mouse brain and observed expression in various areas, especially the isocortex, hippocampus, and cerebellum. We detected TMEM240 in most cerebellar neurons (especially Purkinje cells) and mossy, parallel, and climbing fibers. Finally, we observed a similar pattern of expression in human cerebellar sections.

## Materials and Methods

**Animals and tissue preparation**

Experiments were performed on two-month-old male C57BL/6 mice (Charles River, Lyon, France). Animals were maintained in standard cages and under conventional laboratory conditions (12-h/12-h light/dark cycle, 22 °C), with ad libitum access to food and water. For immunohistochemistry experiments, mice were anesthetized with ketamine (100 mg/kg) and xylazine (10 mg/kg) before transcardial perfusion with 0.9% NaCl and then 4% paraformaldehyde in 0.1 mol/L phosphate-buffered saline (PBS) (pH 7.4). For electron microscopy, 0.1% glutaraldehyde [25%] (for the LR White-embedding method) or 1% glutaraldehyde [25%] and picric acid (for the Araldite embedding method) were added to the perfusion buffer. Brains were removed, post-fixed in 4% paraformaldehyde for 24 h at 4 °C, soaked in 30% sucrose in PBS (for cryoprotection), and then frozen. Sagittal and coronal sections (14 to 40 μm thick) were prepared on a cryostat and mounted on SuperFrost Plus™ Adhesion slides (Thermo Scientific, Brunswick, Germany). Prior to biochemistry analyses, brains were quickly isolated after decapitation and snap-frozen at − 80 °C. The mice were maintained in compliance with European standards on the care and use of laboratory animals. Experimental protocols were approved by the local animal ethical committee (approval APAFIS#2264-2015101320441671 from CEEA75, Lille, France).

**Post-mortem brain samples**

Human brains were obtained from the Lille Neurobank (CRB/CIC1403 Biobank, BB-0033-00030, agreement DC-2008-642), which fulfills the criteria of the local laws and regulations on biological resources with donor consent, data protection, and ethical committee review. We used samples from the cerebellum (superior, ventral, and inferior vermis) from four donors with no signs of motor or cognitive decline. The age at death of the four donors was 22, 25, 37, and 78 years of age. All donors died from cardiovascular disease, with no neurological issues. The post-mortem interval was < 48 h. Neuropathological examination excluded any neurodegenerative diseases and the Braak stage was 0. Formalin-fixed and paraffin-embedded tissues were used for immunofluorescence staining of 7-μm-thick sections.

**Generation of a new rabbit polyclonal antibody recognizing TMEM240**

Rabbit polyclonal antibodies were raised against a synthetic peptide (PYDGDQSVVDASENY) conjugated to keyhole limpet hemocyanin (KLH) (GeneScript, Piscataway, NJ) and corresponding to amino acids 63 to 77 of human TMEM240. Two rabbits were immunized with the purified immunogenic peptide by intramuscular injection at days 0, 14, 28, and 42 using a commercial service (Agro-Bio, La Ferté Saint-Aubin, France). We screened the sera of the two rabbits at days 28 and 42 by ELISA, followed by western blotting and immunohistochemistry. In each experiment, we used mouse cerebellum lysate as a positive control and incubated it with the sera alone or with either the antigenic peptide or recombinant human TMEM240 protein (Origene, Herford, Germany). After screening at days 28 and 42, both rabbits were selected and received a final injection at day 64 before collecting the blood. Only the polyclonal antibody from the rabbit giving the best results during screening was used in this study.

**Immunofluorescence**

The specificity of the TMEM240 antibody (C1orf70, G16, sc-245675, Santa Cruz Biotechnology and TMEM240/63-77) was verified in two competition studies: incubation of the primary antibody (5 μg/mL) with (1) human recombinant protein (recombinant protein of human chromosome 1 open reading frame 70 (C1orf70), #TP325178, Origene, Rockville, MD) or (2) blocking peptide (sc-245675P, Santa Cruz Biotechnology) (25 μg/mL) or TMEM240/63-77 peptide (amino acid sequence: RHHIHYVIPYDGDQ, corresponding to amino acids 55 to 68 of human TMEM240) (25 μg/mL). Another competition study was performed by incubating the TMEM240 antibody with a non-respective blocking peptide (25 μg/mL). Finally, a nonspecific blocking peptide control experiment was performed by incubating both TMEM240 antibodies with a non-competitive BDNF peptide (amino acids 166 to 178 #ANT-010, Alomone Labs, Jerusalem, Israel). Sections were blocked in PBS supplemented with 1% serum from the corresponding antibody species (Table S1). For single and double immunohistochemical staining, 40-µm floating sections were incubated with primary antibodies overnight or up to two days at 4 °C in specific blocking serum. The antibodies used for immunostaining are listed in Table S1. After removal of the primary antibody (by washing with 0.1 mol/L PBS, pH 7.4, supplemented with 0.1% Triton™ X100 (Sigma), slices were incubated with the corresponding secondary fluorescent antibody at room temperature for 45 min. Washing was performed using 0.1 mol/L PBS, pH 7.4. An autofluorescence eliminating reagent (Sudan Black) was used on all sections. The cell nuclei were counterstained with DAPI (Life Technologies, dilution 1/5000) for 5 min. Sections were mounted on SuperFrost Plus™ Adhesion slides (Thermo Scientific, Brunswick, Germany) with KPL mounting medium (Seracare, Life Sciences, Millford, MA, USA).

**Image acquisition**

Fluorescence was observed with Zeiss Axio Imager Z2, confocal Zeiss LSM 710, and Zeiss spinning-disk microscopes. Images were acquired using Zen Lite software and analyzed using Fiji-ImageJ software. The acquisition parameters were selected by the software for each experiment, with corresponding brain sections incubated with the secondary antibody alone. Each image was obtained using the selected acquisition parameters to avoid background staining. For whole brain and cerebellum images, tiles and maximum intensity projection images were obtained on 35-µm-thick sections. Relative grayscale values were obtained by manually drawing regions of interest (ROIs) for various brain structures, such as the cerebellum, deep nuclei, and layers of the cerebellar cortex. Pixel intensities in each ROI were quantified on RGB images converted into grayscale images. The background values were subtracted prior to quantification (the threshold was the same within experiments). The data were normalized by square root transformation to obtain a normal distribution and reduce skewness. Finally, all values were divided by the ROI surface to obtain the relative grayscale values. At least three independent experiments were performed before quantification or colocalization analyses. For statistical analysis, at least 10 images were measured for three different animals for brain structure quantification. Thus, there were 30 independent measurements for each structure. Co-localization of staining was evaluated with Pearson's correlation coefficient (PCC) using Zen (Zen Black, Carl Zeiss Microscopy, GmbH, Munich, Germany) or Imaris software (Bitplane AG, Zurich, Switzerland).

**Preparation of Pre- and Post-synaptic Density Fractions**

Cerebellar lysates were homogenized in Tris-sucrose buffer (10 mM Tris, 10% sucrose) and the protein concentrations determined using the BCA assay. In total, 100 µg protein lysate was homogenized in 500 µL Tris-sucrose buffer. Homogenates were sonicated and centrifuged at 1000g at 4 °C for 10 min. Supernatants were collected and centrifuged at 12,000g at 4 °C for 20 min. Pellets were washed in 500 µL Tris-EDTA buffer (4 mM Tris, 1 mM EDTA, pH 7.4) and centrifuged two times at 12,000g at 4 °C for 20 min. Pellets were resuspended in Tris-NaCl Triton buffer (20 mM Tris, 100 mM NaCl, 0.5% Triton™X100, pH 7.2) and homogenized at 4 °C for 15 min. After centrifugation at 15,000g at 4 °C for 20 min, the supernatants, corresponding to post-synaptic density (PSD) fractions, were washed three times in Tris-NaCl Triton buffer. The pellets, corresponding to non-PSD fractions, were washed in 100 µL Tris-NaCl-Triton. Samples were mixed with NuPage® LDS 2× sample buffer supplemented with 20% NuPAGE® sample reducing agents (Invitrogen, Carlsbad, CA, USA) and equal volumes were loaded onto gels and analyzed by western blotting. Duplicate samples for PSD and non-PSD fractions were loaded.

**Western blotting**

Samples were homogenized in an appropriate volume of Tris-sucrose buffer and protein concentrations determined using the BCA assay. One volume of NuPage® LDS 2× sample buffer, supplemented with 20% NuPAGE® sample reducing agents (Invitrogen, Carlsbad, CA, USA), was added to the homogenates. Samples were then heated 10 min at 100 °C. After loading 30 µg total protein lysate, molecular-weight markers (Novex and Magic Marks, Life Technologies, Carlsbad, CA, USA), and 10 ng recombinant human protein (recombinant protein of human chromosome 1 open reading frame 70 (C1orf70), #TP325178, Origene, Rockville, MD) onto precast 12% Criterion XT Bis-Tris polyacrylamide gels (Bio-Rad, Hercules, CA, USA), electrophoresis was carried out by applying 165 V for 45 min in NuPAGE® MES SDS running buffer (1×). Proteins were transferred to a 0.2-µM nitrocellulose membrane (G&E Healthcare, Chicago, IL, USA) at 100 V for 40 min. The quality of the protein transfer was determined by reversible Ponceau Red coloration (0.2% Xylidine Ponceau Red and 3% trichloroacetic acid). Membranes were blocked in 25 mM Tris-HCl pH 8.0, 150 mM NaCl, 0.1% Tween-20 (v/v) (TBS-T) supplemented with 5% (w/v) skimmed milk (TBS-M) or 5% (w/v) bovine serum albumin (TBS-BSA), depending on the antibody, for 1 h (Table S2). Membranes were washed in TBS-T before incubation with primary antibodies overnight at 4 °C. After washing in TBST, membranes were incubated with secondary antibodies for 45 min at room temperature. Revelation was performed using ECL™ Western Blotting Detection Reagents (G&E Healthcare, Chicago, IL, USA).

**Immunogold electron microscopy**

Two different fixation procedures were used for the electron microscopy analysis: LR White embedding gives higher sensitivity, whereas Araldite embedding preserves cellular structures. The cerebellum was rapidly dissected, incubated in fixation buffer at 4 °C, and post-fixed in 1% osmic acid, 0.1 M phosphate buffer. The sections were dehydrated using a series of increasing ethanol concentrations, impregnated with 96% LR White resin/ethanol or Araldite/absolute ethanol, and then stored in pure buffer at 4 °C. After polymerization (4 °C under UV for LR White or 56 °C for Araldite), the sections (85-nm ultra-thin sections) were incubated with the anti-TMEM240 antibody (Santa Cruz Biotechnology) for three days in 0.1 M Tris, 0.15 M NaCl, 1% BSA, pH 7.4, and 1% normal donkey serum, supplemented (for Araldite embedding) with 30% $H_2O$. The same buffer was used for washing. The sections were then incubated with donkey anti-goat antibodies conjugated to 12-nm gold particles (in 0.1 M Tris, 0.5 M NaCl, 1% BSA, pH 7.4, and normal donkey serum 1%) for 1 h. After washing with the same buffer, the contrast was developed by exposure to 2% uranyl acetate in 50% ethanol for 10 s (LR White embedding) or 2% uranyl acetate and then lead citrate for 8 min each (Araldite embedding). The sections were visualized under a Zeiss EM 900 microscope at magnifications from $\times 3000$ to $\times 140,000$.

**Statistical analysis**

Data are reported as the mean ± standard error of the mean (SEM). All statistical analyses were performed using GraphPad Prism software version 7 (GraphPad Software Inc., La Jolla, CA, USA). The normality of the data distribution was assessed by the Shapiro-Wilk test. Differences between mean values were determined using the paired Student's t test for data when comparing an independent variable between two groups. For the analysis of one independent variable (mean relative grayscale value in the whole structure of interest) between more than two groups (mean relative grayscale value in different parts of a structure of interest), a one-way analysis of variances (ANOVA) was used with a post hoc Fisher's least significant difference (LSD) test. The threshold for statistical significance was set to $p < 0.05$.

## Results

*TMEM240 antibodies specificity*

Competition studies performed in the anterior lobe verified the specificity of the TMEM240 antibodies (Figures S1 and S2 for the TMEM240 sc-245675 (Santa Cruz Biotechnology) and TMEM240/63-77 antibodies, respectively). Incubation of the TMEM240 antibodies with the respective immunogenic blocking peptide (Figures S1.b1-b2 and S2.b1-b2) or human recombinant protein (Figures S1.c1-c2 and S2.c1-c2) abolished the TMEM240 immunoreactivity observed when incubating thin sections with the TMEM240 antibodies alone (Figures S1.a1-a2 and S2.a1-a2). Incubation of the TMEM240 antibodies with a non-competitive BDNF peptide (Figures S1.d1-d2 and S2.d1-d2) or a non-respective immunogenic blocking peptide (Data not shown) did not abolish TMEM240 immunoreactivity.

*TMEM240 protein localization in the mouse brain*

TMEM240 immunoreactivity was quantified by immunofluorescence labeling of TMEM240 with Santa Cruz Biotechnology (Fig. 1a) and TMEM240/63-77 (Fig. 1b) antibodies on sagittal brain sections. We observed no staining of the corpus callosum. TMEM240 immunoreactivity (Santa Cruz Biotechnology antibody) was lower in the striatum ($5.52 \pm 0.1$ U, $p < 0.0001$) and hindbrain ($5.54 \pm 0.1$ U, $p < 0.0001$) than in the brain as a whole ($6.09 \pm 0.07$ U). TMEM240 immunoreactivity in the midbrain showed mean values similar to those obtained in the brain as a whole ($6.16 \pm 0.19$ U, $p = 0.98$) and interbrain (defined as the brain region encompassing the thalamus and hypothalamus ($6.29 \pm 0.15$ U, $p = 0.13$)). We detected higher values in the isocortex ($7.95 \pm 0.17$ U, $p < 0.0001$), hippocampus ($7.39 \pm 0.13$ U, $p < 0.0001$), and cerebellum ($7.30 \pm 0.35$ U, $p < 0.0001$) (Fig. 1c). We also assessed TMEM240 immunoreactivity in the dentate gyrus, caudate nucleus, putamen, cerebral cortex, and cerebellum, as databases show that human TMEM240 is highly expressed in these regions [8]. We observed higher values in the CA1 and CA2 pyramidal layers ($9.25 \pm 0.07$ U and $8.78 \pm 0.12$ U, respectively, $p < 0.0001$) and granular layer of the dentate gyrus ($8.40 \pm 0.05$ U, $p < 0.0001$) than in the hippocampal formation as a whole ($7.39 \pm 0.03$ U). The lowest value was obtained in the CA3 field ($6.68 \pm 0.08$ U, $p < 0.0001$) (Figure S3.a). In the area of the striatum, we visualized higher expression in the nucleus accumbens than the caudoputamen ($6.48 \pm 0.3$ U vs $5.36 \pm 0.05$ U, respectively, $p < 0.0001$) (Figure S3.b). Compared with the mean intensity in the isocortex as a whole ($7.95 \pm 0.17$ U), we observed the lowest value in the orbital area ($6.74 \pm 0.14$ U, $p < 0.0001$), values close to that of the isocortex as a whole in the somatosensory-related area ($7.7 \pm 0.04$ U), somatomotor-related area ($7.78 \pm 0.09$ U), visual areas ($7.86 \pm 0.04$ U), and posterior parietal association areas (7.78 ± 0.05 U), and the highest value in the retrosplenial area (8.21 ± 0.03 U, $p < 0.0001$) (Figure S3.c). We obtained similar TMEM240 quantification values with the TMEM240/63-77 antibody (Figs. 1d, and S3.d-f). We observed a marked background with the commercial TMEM240 antibody (Santa Cruz Biotechnology). However, our home-made antibody (TMEM240/63-77) allowed us to obtain similar staining in the brain and cerebellum with less background. The quantification of TMEM240 expression in the regions, areas, and subparts of the adult mouse brain is presented in Table S3.

*TMEM240 protein expression in the cerebellum*

We analyzed TMEM240 immunoreactivity on sagittal, parasagittal, and coronal cerebellar sections in each cerebellar lobule, cerebellar fiber tracts, and the fastigial, interposed, and dentate cerebellar nuclei. We detected immunoreactivity for TMEM240 in all cerebellar structures analyzed and mild differential expressions between lobules with both TMEM240 antibodies, but with lower values in the cerebellar fiber tracts (Figs. 2 and S4). In the cerebellar cortex (sagittal sections, TMEM240 Santa Cruz Biotechnology antibody, Fig. 2a), we observed a lower value in the lingula (lobule I) (6.65 ± 0.24 U, $p < 0.0001$) and higher values in the culmen (lobules IV–V) (7.64 ± 0.27 U, $p < 0.0001$) and folium-tuber vermis (lobule VII) (7.41 ± 0.24 U, $p < 0.0001$) (Fig. 2c) than the mean intensity value for the cerebellum as a whole (7.24 ± 0.60 U). We obtained similar TMEM240 quantification values with the TMEM240/63-77 antibody (Fig. 2b, d). We detected TMEM240 in the fastigial nucleus (5.26 ± 0.32 U) and interposed nucleus (5.71 ± 0.32 U) (Fig. 2c). In the cerebellar cortex (parasagittal sections, TMEM240 Santa Cruz Biotechnology antibody, Figures S4.a-b), we detected lower values in the ansiform Crus 2 lobule (6.70 ± 0.50 U, $p = 0.0042$) and paramedian lobule (6.82 ± 0.46 U, $p = 0.13$) than the mean intensity value for the cerebellum as a whole (7.12 ± 0.42 U). We observed higher values in the simple lobule (7.62 ± 0.42 U, $p = 0.0002$) and ansiform Crus 1 lobule (7.60 ± 0.49 U, $p = 0.0006$). We observed the highest value in the copula pyramidis lobule (7.75 ± 0.49 U, $p < 0.0001$). We also detected TMEM240 in the dentate nucleus (7.03 ± 0.38 U) (Figures S4.a-b). We obtained similar staining in the parasagittal sections with the TMEM240/63-77 antibody (Figure S4.c). In the cerebellar nuclei, TMEM240 was highly expressed in neurons with large cellular bodies and showed a punctuate pattern (Figure S4.d). We detected TMEM240 in the molecular and granular layers and major immunoreactivity in the Purkinje-cell layers with both antibodies, as shown by low magnification (Fig. 3a, d). In the molecular layer, punctuate TMEM240 expression was detected in the Purkinje-cell membrane, the soma, and dendrites (Fig. 3b, c, e, f).

*Cellular distribution of TMEM240*

TMEM240 immunoreactivity (Santa Cruz Biotechnology) co-localized strongly (PCC > 0.6) with immunostaining for calbindin, a specific Purkinje-cell marker (Fig. 4a1–a3). At a higher magnification, we found immunostaining for TMEM240 in the Purkinje-cell membrane, soma, axons, and dendrites, but not in the nuclei (Fig. 4b1–b3). Similar co-staining with calbindin was obtained with the TMEM240/63-77 antibody (Figure S5). Interestingly, immunoreactivity for TMEM240 was found in several calbindin-negative cells in the Purkinje-cell and molecular layers, suggesting that TMEM240 is expressed in stellate and basket cells (Fig. 4b3). Costaining with parvalbumin (a molecular layer interneuron marker) and TMEM240 (Santa Cruz Biotechnology) showed positive cells for both antibodies. These cells may thus correspond to basket cells in the Purkinje-cell layer and stellate cells in the outer two-thirds of the molecular layer (Figure S6). Moreover, certain cells in the molecular layer were positive for TMEM240 only, which may correspond to excitatory neurons in the molecular layer (Figure S6). The immunoreactivity of TMEM240 in glial cells was negative, as demonstrated by co-staining with glial fibrillary acidic protein (data not shown). We also co-stained for the vesicular glutamate transporters VGLUT1 (Fig. 5a1–a5) and VGLUT2 (Fig. 5b1–b5) to examine the localization of TMEM240 in the cerebellar connections. Co-immunostaining with VGLUT1 in the molecular layer (Fig. 5a4) showed TMEM240 to also be localized to the parallel fibers. In the granular layer, co-staining showed TMEM240 to be localized to the glomeruli, with partial co-staining of TMEM240 and VGLUT1, which could correspond to mossy fibers (Fig. 5a5). Co-immunostaining with VGLUT2 in the molecular layer (Fig. 5b4) showed TMEM240 localization to the climbing fibers. In the glomeruli, coimmunostaining with VGLUT2 showed TMEM240 localization to climbing and mossy fibers (Fig. 5b5).

*Synaptic distribution of TMEM240 immunoreactivity in the murine and human brains*

We co-stained for TMEM240 (Santa Cruz Biotechnology) and synaptophysin to examine TMEM240 localization at the cerebellar synapses (Fig. 6a–d). Intense co-immunostaining (PCC > 0.6) showed the presence of TMEM240 in all synapses in all three layers of the cerebellar cortex. At a higher magnification, we observed co-immunostaining with synaptophysin in synapses connecting to the membrane of a Purkinje-cell soma (Fig. 6d). PSD fractionation showed TMEM240 localization in enriched-post-synaptic fractions (Fig. 6e) with both TMEM240 (Santa Cruz Biotechnology) and TMEM240/63-77 antibodies. We performed electron microscopy on TMEM240 immunogold-labeled cerebellar sections in mice (Santa Cruz Biotechnology). Araldite embedding (Fig. 7a, b) and LR White embedding (Fig. 7c, d) showed TMEM240 to be localized to synapses connecting to the membrane Purkinje-cell soma on the post-synaptic side of the Purkinje-cell.

Finally, we examined the localization of TMEM240 in adult human cerebellar cortex sections with both TMEM240 antibodies. As in the mouse cerebellum (Fig. 6), we detected TMEM240 immunoreactivity in each layer of the human cerebellar cortex (Fig. 8.a1, b1-b3). Co-immunostaining with synaptophysin (PCC > 0.6) showed a similar localization between the the mouse (Figure 6) and Human in the glomeruli (Figure 8.a2), molecular layer (Figure 8.a3) and Purkinje cells (Figure 8.a4).

**Discussion**

Recently, a study by Seki *et al.* showed endogenous TMEM240 expression in the mouse brain by immunoblotting; the protein was broadly expressed in the cerebral cortex, hippocampus, striatum, thalamus, hypothalamus, midbrain, and, in particular, the cerebellum [18]. Our present immunofluorescence results are in accordance with the findings of Seki *et al.* and the data on mRNA expression in human brain databases [8]. We found TMEM240 expression to be predominant in the cerebellar cortex and dentate nucleus. Most SCA21 patients present with gait and limb ataxia and cerebellar symptoms (such as saccadic pursuit or gaze-evoked nystagmus, accompanied by brainstem oculomotor disturbances, resulting in slow ocular saccades). However, SCA21 has been recently characterized as a multisystem neurodegenerative disease, with extracerebellar oculomotor disturbances and hypo/hyperkinetic movement disorders [6]. These symptoms may be explained by our observation of TMEM240 expression in the cerebral cortex and striatum. Although cognitive impairments (such as CCAS) may be associated with cerebellar dysfunction [19], memory disorders, impaired executive function, frontal behavior disorders, and mental retardation are more likely to reflect hippocampus and/or cortical dysfunction in SCA21 patients [20]. Focused, in-depth studies of TMEM240 expression in other parts of the brain (notably the hippocampus, cerebral cortex, and striatum) would be of great interest.

During our study, we focused on male mice. This is a limitation as TMEM240 could have some gender-specific differential expression. However, in our previous clinical publications [3, 20], we did not observe any gender-based difference within the SCA21 patients, assuming that impact of TMEM240 gene mutations, and presumable TMEM240 protein function, is not gender-related.

Our study characterized TMEM240 expression in the cerebellar neuronal network. Immunofluorescence experiments showed TMEM240 staining in the neurons of the cerebellar cortex and in numerous neuronal projections. TMEM240 was present in afferents from various structures, particularly in neurons that connect to Purkinje cells. We also observed the presence of TMEM240 protein in a high proportion of synapses in the cerebellar cortex; this finding is in accordance with that of a previous report of TMEM240 protein in murine synaptosomes [7].

TMEM240 is widely expressed in Purkinje-cell dendrites, bodies, and axons. Progressive cerebellar cell degeneration and neuronal loss are major, frequent features of SCA. In most cases, neuroimaging and histological analyses have identified Purkinje-cell death as the main pathological feature [21]. Neuropathological data from an autopsied patient with SCA21 revealed a severe loss of Purkinje cells [5]. Our observation of predominant TMEM240 expression in Purkinje cells is consistent with neuropathological evidence showing that the neurodegeneration in SCA21 primarily affects these neurons. TMEM240 immunostaining of Purkinje cells from normal mice had a punctuate pattern and was distributed equally throughout the soma. This is in contradiction to the report of Seki et al. on cerebellar primary cultures, in which TMEM240 expression was observed in large, cytoplasmic vesicles in the Purkinje-cell soma, dendrites, and axons [20]. The discrepancy between our results and those of their ex vivo cell-based model may be due to overexpression of a flagged version of the TMEM240 protein. The vesicular glutamate transporters of cerebellar afferents mediate glutamate uptake into the synaptic vesicles of excitatory neurons [22]. In the molecular layer, VGLUT2 is exclusively expressed in climbing fibers, whereas parallel fibers express VGLUT1 for vesicular glutamate uptake. In the granular layer, both transporters are detected in mossy fiber terminals [23]. Co-immunostaining with VGLUT1 and VGLUT2 demonstrated the presence of TMEM240 in glutamatergic terminals from climbing, parallel, and mossy fibers. Mutations in glutamate transporters and/or receptors lead to Purkinje-cell degeneration and ataxia in murine models and humans [24, 25]. TMEM240 mutations in SCA21 may interfere with neuronal glutamate homeostasis and thus cerebellar function. Functional studies of wild type and mutated proteins in cell-based and animal models are needed to validate this hypothesis.

**Conclusion**

We characterized TMEM240 expression in the cerebellar neuronal network, including neurons of the cerebellar cortex, particularly Purkinje cells. Our work provides insights into the physiological expression of TMEM240 in the cerebellum. These insights are likely to be of great value in the establishment of murine or zebrafish models of SCA21.

## Compliance with Ethical Standards

**Ethical approval**

All procedures performed in studies involving animals were in accordance with the ethical standards of the institution at which the studies were conducted. Experimental protocols were approved by the local animal ethical committee (approval APAFIS#2264-2015101320441671 from CEEA75, Lille, France). Human brains were obtained from the Lille Neurobank (CRB/CIC1403 Biobank, BB-0033-00030, agreement DC-2008-642), which fulfills the criteria of the local laws and regulations on biological resources with donor consent, data protection, and ethical committee review.

**Disclosure of potential conflict of interest**

The authors disclose no conflicts of interest.

**Consent for publication**

All authors reviewed the current manuscript and approved it for submission.

**Availability of data and material**

All data generated or analyzed during this study are included in this published article and its supplementary information files.

## Acknowledgments

We thank the Lille NeuroBank (CHRU-Lille) for providing the human brain tissue, and Nicolas Van Poucke from Lille Biology and Pathology center. We also acknowledge Dr Meryem Tardivel and Antonino Bongiovanni from cellular imagery platform for confocal microscopy experiments. Lastly, we thank Dr Khalid Hamid El Hachimi (ICM, Paris) for his precious advices on interpreting the electron microscopy data.

**Funding**

The research leading to these results was funded by the French government's LabEx program ("Development of Innovative Strategies for a Transdisciplinary approach to Alzheimer's disease"- DISTALZ), the University of Lille, the Institut National de la Santé et de la Recherche Médicale (INSERM), and the *Connaître les Syndrômes Cérébelleux* (CSC) charity.

**Figure legends**

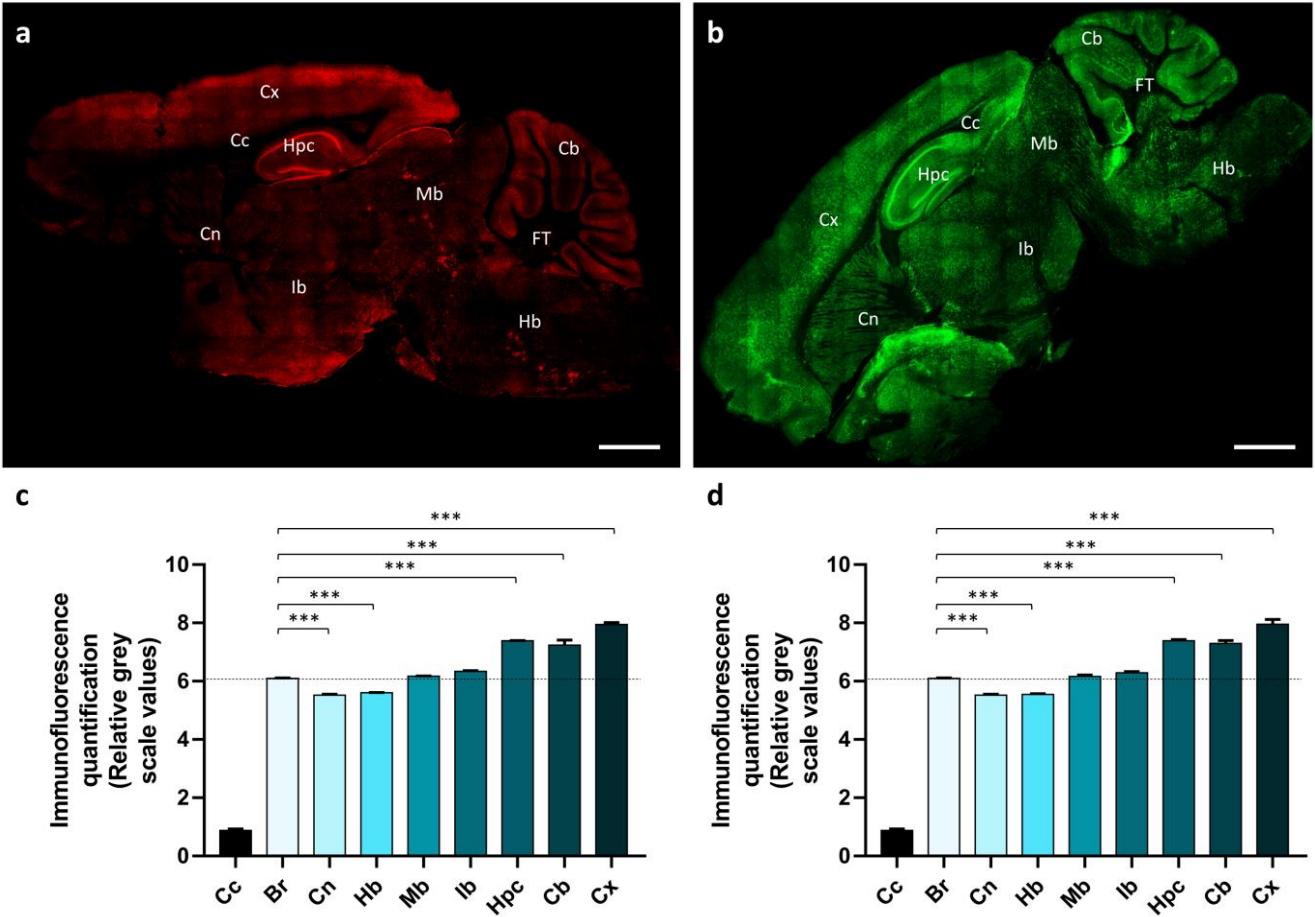

**Fig. 1** TMEM240 expression in the mouse brain.

TMEM240 immunoreactivity was detected by immunofluorescence labeling of sagittal brain sections. Staining with the TMEM240 antibody from Santa Cruz Biotechnology is shown in red (a) and that with TMEM240/63-77 antibody in green (b). Scale bar, 1000 μm. Immunofluorescence quantification of TMEM240 immunoreactivity in several brain areas expressed as the number of pixels normalized to the relative grayscale values. The total brain area (Br) was used as the reference for the statistical analysis using the TMEM240 antibody from Santa Cruz Biotechnology (c) and the TMEM240/63-77 antibody (d). $*p < 0.05$, $**p < 0.01$, $***p < 0.001$. Images were obtained with a Zeiss spinning-disk confocal microscope. Cb cerebellum, Cc corpus callosum, Cn cerebral nuclei, Cx isocortex, FT cerebellar fiber tracts, Hb hindbrain, Hpc hippocampal formation, Ib interbrain, Mb midbrain

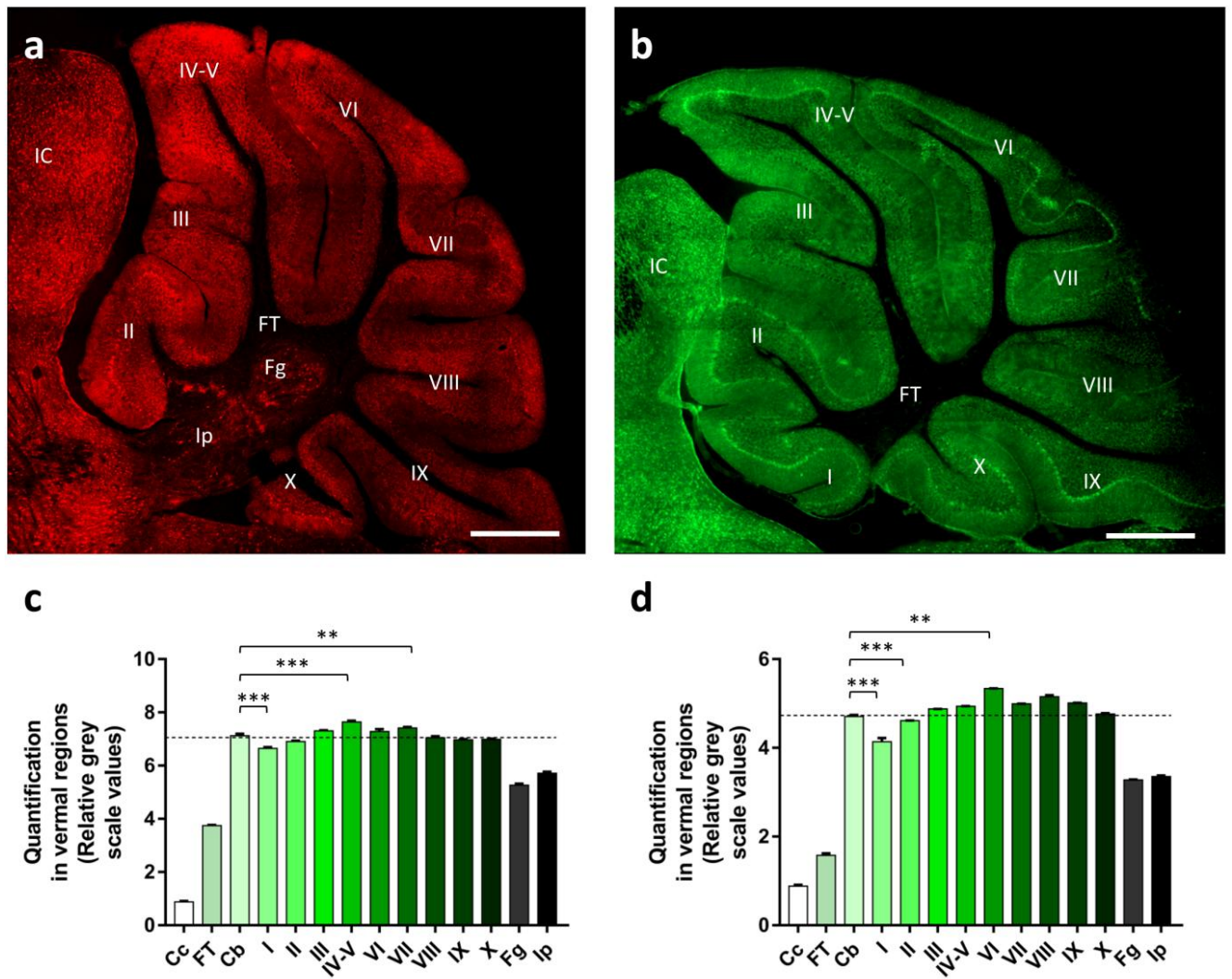

**Fig. 2** TMEM240 expression in the mouse cerebellum.

TMEM240 immunoreactivity was detected by immunofluorescence labeling of sagittal cerebellar sections. Staining with the TMEM240 antibody from Santa Cruz Biotechnology is shown in red (a) and that with the TMEM240/63-77 antibody in green (b). Scale bar, 500 μm. Immunofluorescence quantification in sagittal cerebellar sections is expressed as the number of pixels normalized to the relative grayscale values using the Santa Cruz Biotechnology antibody (c) or TMEM240/63-77 antibody (d). Total cerebellar staining (Cb) was used as a reference measurement for statistical analysis. *$p < 0.05$, **$p < 0.01$, ***$p < 0.001$. Images were obtained by confocal microscopy. Cc corpus callosum, Fg fastigial nucleus, FT cerebellar fiber tracts, IC inferior colliculus, Ip interposed nucleus, I lingula, II central lobule II, III central lobule III, IV-V culmen (lobules IV-V), VI declive (lobule VI), VII folium-tuber vermis (lobule VII), VIII pyramus (lobule VIII), IX uvula (lobule IX), X nodulus (lobule X)

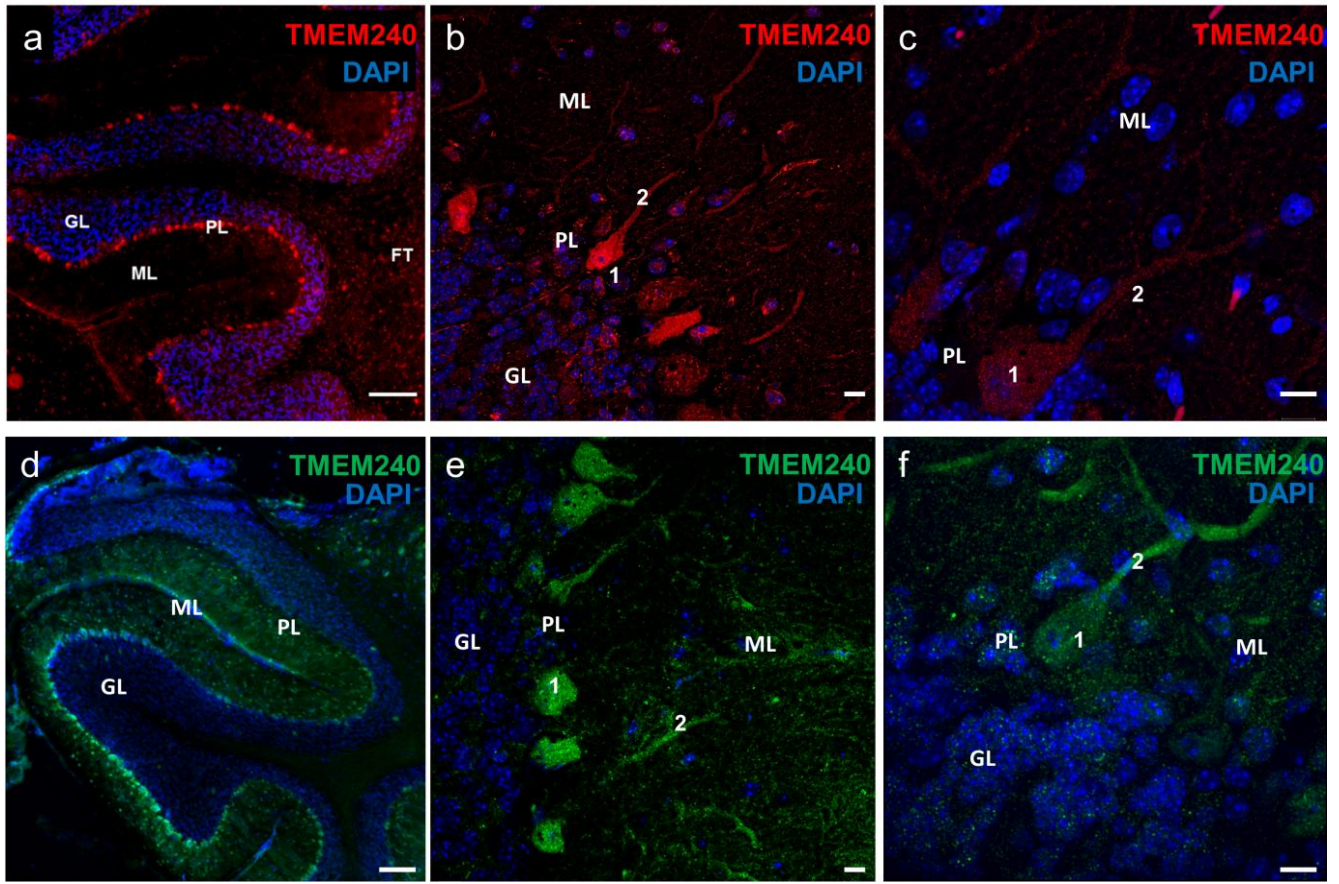

**Fig. 3** Cellular expression of TMEM240 in the mouse cerebellar cortex.

(a) Immunofluorescence labeling of nuclear DNA using DAPI (blue) and TMEM240 antibody (TMEM240 antibody from Santa Cruz Biotechnology in red) in sagittal cerebellar sections. Scale bar, 100 μm. (b) Magnification of TMEM240 staining in the cerebellar cortex (Santa Cruz Biotechnology, in red). Scale bar, 10 μm. (c) Magnification of punctuate TMEM240 staining in Purkinje-cell bodies and dendrites. Scale bar, 10 μm. (d) Immunofluorescence labeling of nuclear DNA by DAPI (blue) and TMEM240 antibody (TMEM240/63-77 antibody in green) in sagittal cerebellar sections. Scale bar, 100 μm. (e) Magnification of TMEM240 staining in the cerebellar cortex (TMEM240/63-77 antibody). Scale bar, 10 μm. (f) Magnification of punctuate TMEM240 staining in Purkinje-cell bodies and dendrites. Scale bar, 10 μm. Images were obtained by confocal microscopy. FT fiber tracts, GL granular layer, ML molecular layer, PL Purkinje-cell layer, PC Purkinje cell. Purkinje bodies (1) and dendrites (2)

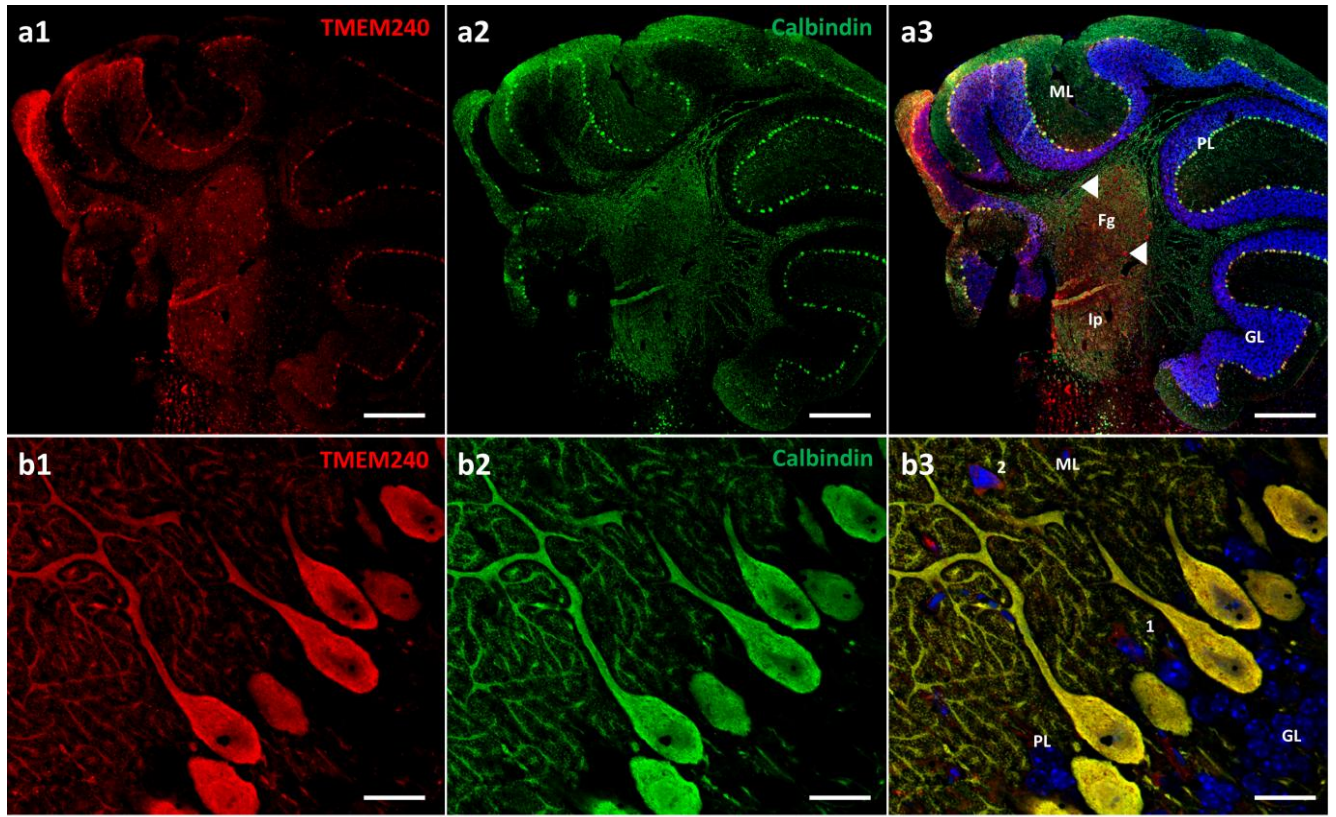

**Fig. 4** Expression of TMEM240 in Purkinje cells.

(a1–a3) Double immunofluorescence labeling of TMEM240 (Santa Cruz Biotechnology) (in red) and calbindin (in green) in sagittal cerebellar sections. Scale bar, 100 μm. White arrows indicate TMEM240-positive cells in white matter and deep nuclei. (b1–b3) Magnification of the double immunofluorescence labeling image of TMEM240 (Santa Cruz Biotechnology) (in red) and calbindin (in green) of a sagittal cerebellar section. TMEM240-positive cells corresponding to the labeling of basket cells (1) and stellate cells (2). Images were obtained by confocal microscopy. Scale bar, 20 μm. Fg fastigial nucleus, GL granular layer, Ip interposed nucleus, ML molecular layer, PL Purkinje-cell layer

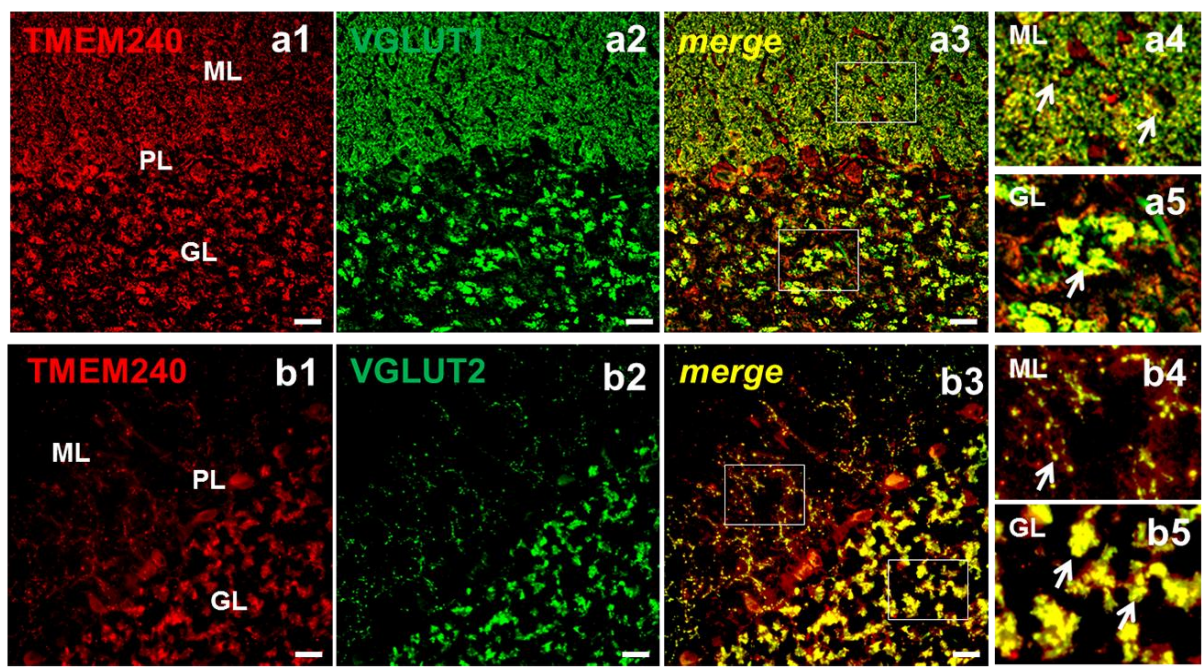

**Fig. 5** Localization of TMEM240 in mossy, parallel and climbing fibers.

(a1–a5) Double immunofluorescence labeling of TMEM240 (Santa Cruz Biotechnology) (in red) and VGLUT1 (in green) in sagittal cerebellar sections. Scale bar, 50 μm. (a4) Magnification of an image showing co-staining in the molecular layer. The white arrows indicate TMEM240 localization in parallel fibers. (a5) Magnification of an image showing co-staining in the granular layer. The white arrow indicates TMEM240 and VGLUT1 co-localization in glomeruli. (b1–b5) Double immunofluorescence labeling of TMEM240 (Santa Cruz Biotechnology) (in red) and VGLUT2 (in green) in sagittal cerebellar sections. Scale bar, 50 μm. (b4) Magnification of an image showing co-staining in the molecular layer. The white arrow indicates TMEM240 localization in climbing fibers. (b5) Magnification of an image showing co-staining in the granular layer. The white arrows indicate TMEM240 in climbing fibers and mossy fibers. Images were obtained by confocal microscopy. GL granular layer, ML molecular layer, PL Purkinje-cell layer

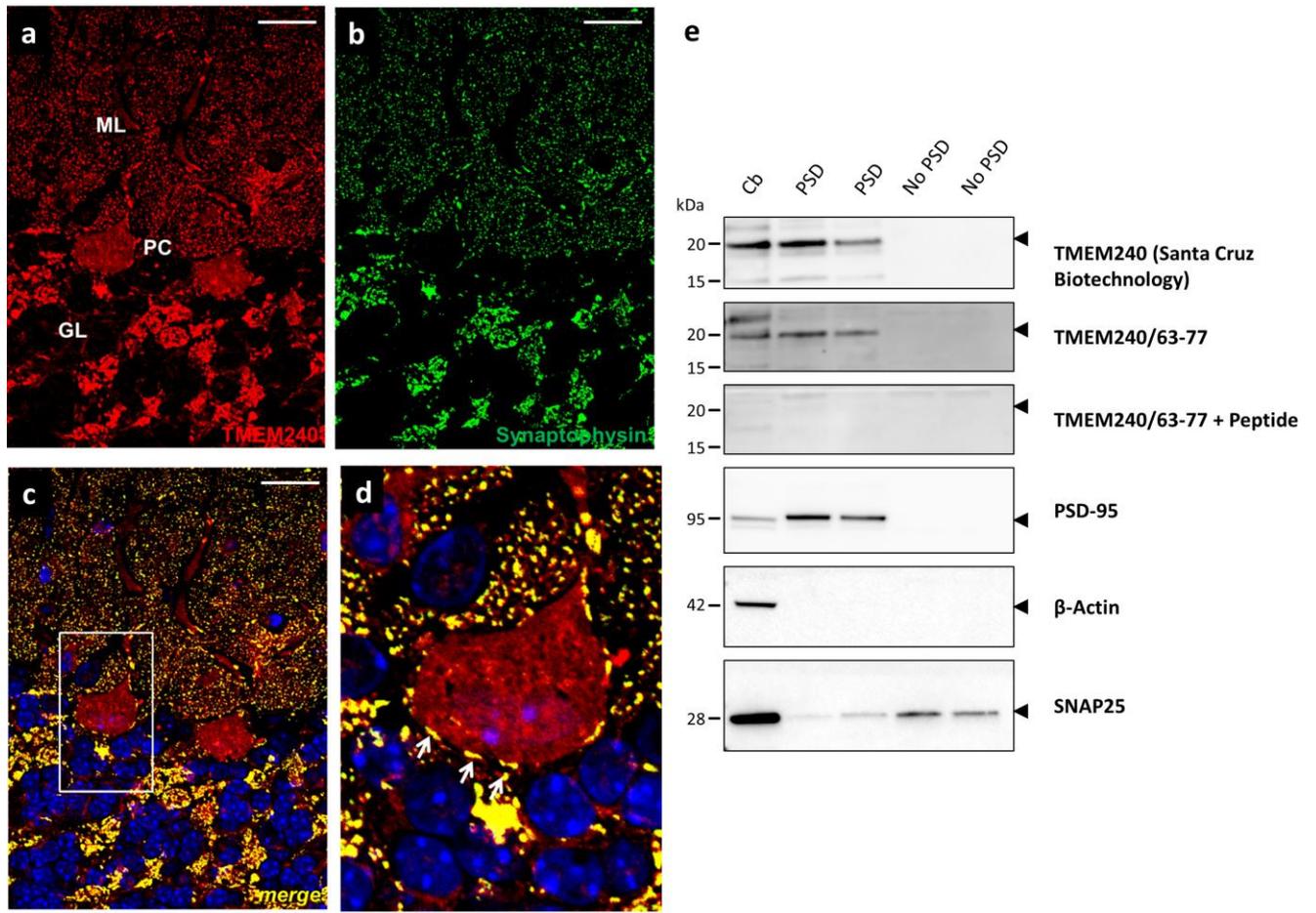

**Fig. 6** TMEM240 localization in synapses.

(a–d) Double immunofluorescence staining of TMEM240 in red (Santa Cruz Biotechnology) (a) and synaptophysin in green (b) in murine sagittal cerebellar sections. (c) Double immunofluorescence (merge) of both TMEM240 and synaptophysin. Scale bar, 10 μm. (d) Magnification of double immunofluorescence staining, focusing on a Purkinje-cell soma. The white arrows indicate TMEM240/synaptophysin coimmunostaining at the Purkinje-cell membrane. Images were obtained by confocal microscopy. (e) Western-blot analysis of TMEM240 localization in cerebellum lysate (Cb), post-synaptic density fractions (PSD), and non-postsynaptic density fractions (No PSD). GL granular layer, ML molecular layer, PC Purkinje cell

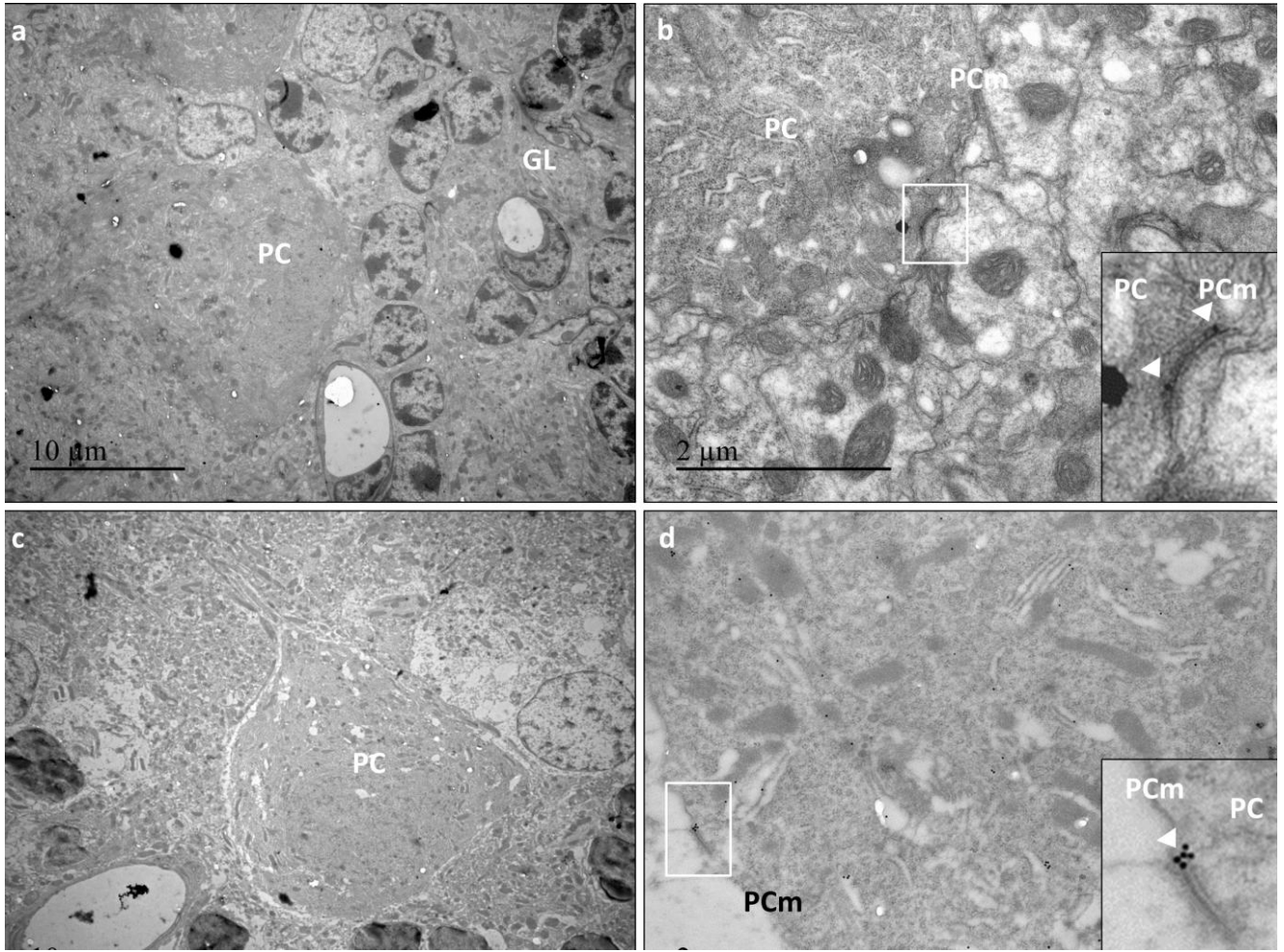

**Fig. 7** Electron microscopy analysis.

Transmission electron micrographs of TMEM240 immunogold staining (Santa Cruz Biotechnology antibody) of murine cerebellar sections embedded in Araldite resin (a, b) or LR White resin (c, d). Micrographs showing Purkinje-cell soma (a, c). Scale bar, 10 μm. Magnification of immunogold stained Purkinje-cell synapses (b, d). Scale bar, 2 μm. The white arrows indicate TMEM240 localization on the post-synaptic side of the Purkinje-cell membrane. GL granular layer, PC Purkinje cell, PCm Purkinje-cell membrane

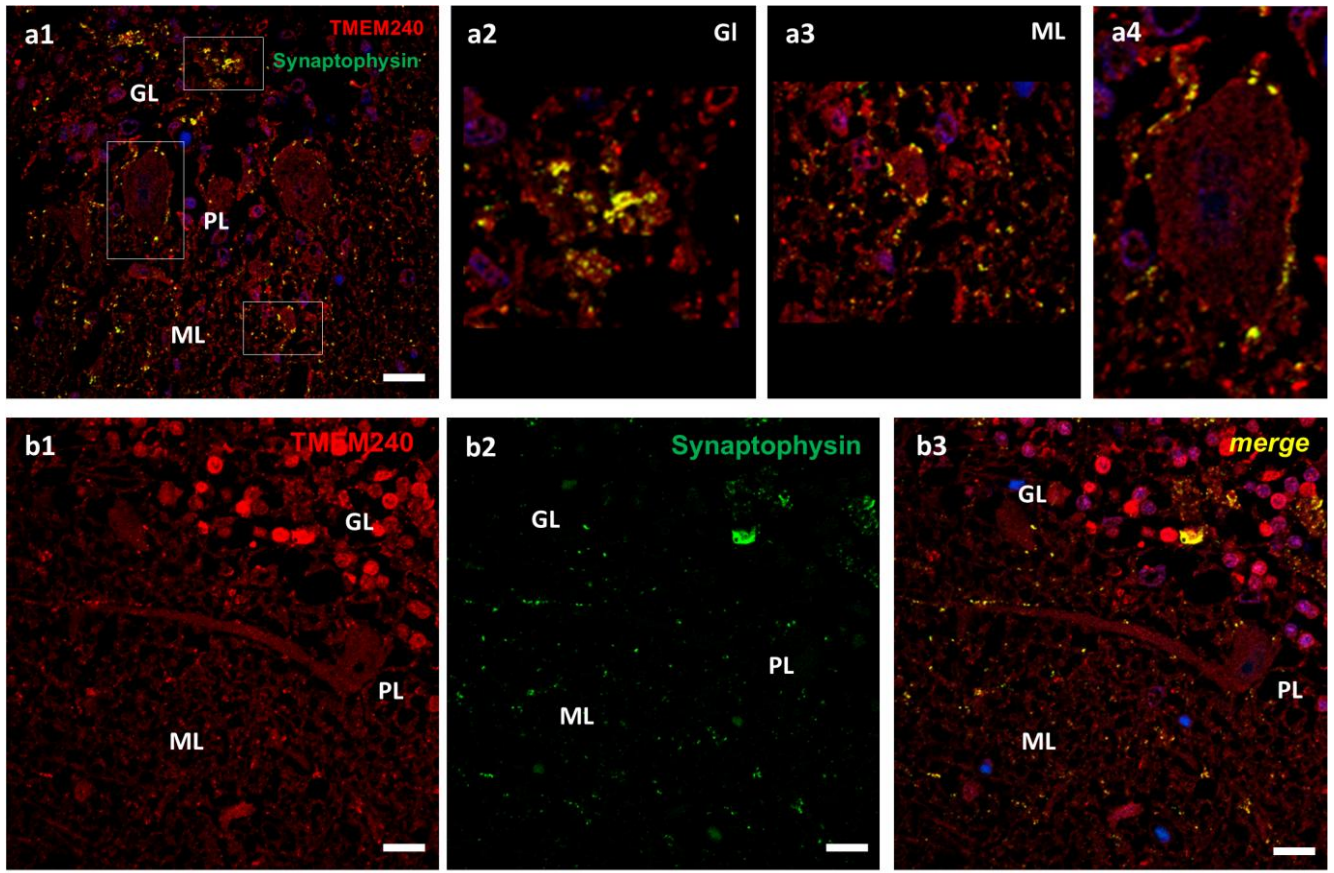

**Fig. 8** TMEM240 localization in human cerebellar cortex sections.

Double immunostaining of TMEM240 (Santa Cruz Biotechnology) (in red) and synaptophysin (in green) on human cerebellar cortex section (a1, b1-b3). Scale bar: 10 µm. Magnifications of a glomerulus (a2), the molecular layer (a3), and a Purkinje-cell soma (a4). Gl: glomerulus, GL: granular layer, ML: molecular layer, PL: Purkinje-cell layer

## Contributors

Mégane HOMA: Had a major role in the acquisition of data, including all immunohistochemical analyses; interpreted the data, and drafted the manuscript.

Anne LOYENS: Had a major role in the acquisition of transmission electron micrographs.

Sabiha EDDARKAOUI: Had a major role in the production of a new polyclonal antibody.

Emilie FAIVRE: Had a major role in animal and tissue preparation. Revised the manuscript for critical intellectual content.

Vincent DERAMECOURT and Claude-Alain MAURAGE: Supply of human cerebellar samples. Selection of images.

Luc BUÉE: Revised the manuscript for critical intellectual content.

Vincent HUIN and Bernard SABLONNIÈRE: Interpreted the data, and revised the manuscript for critical intellectual content.

## Supplementary material

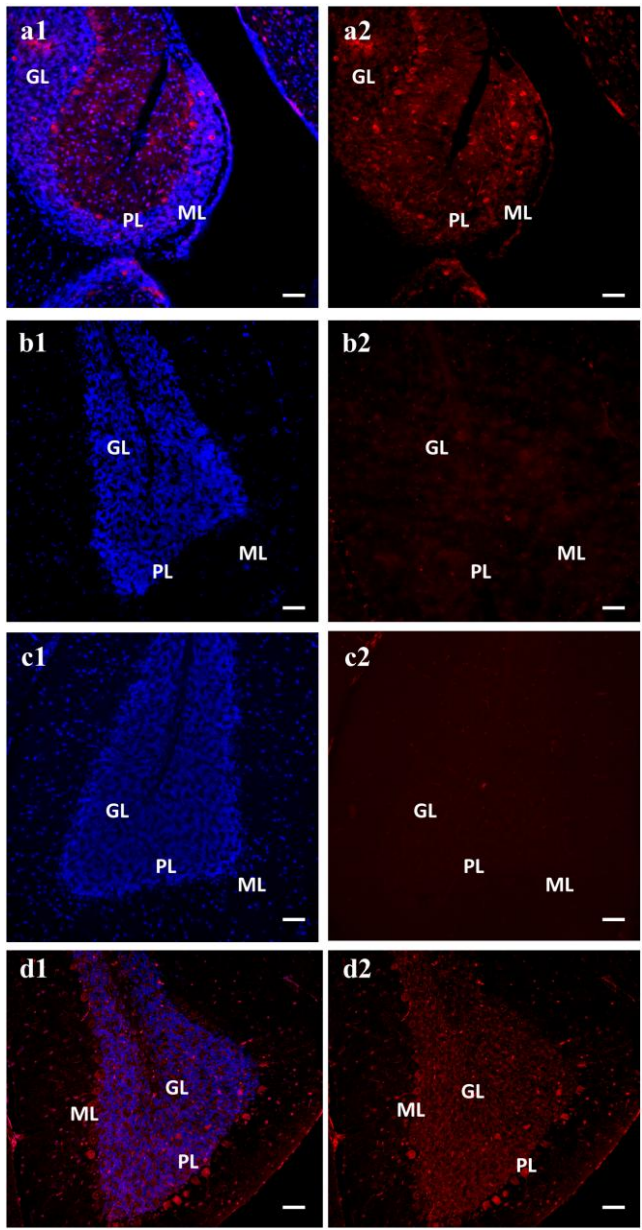

**Supplemental Figure S1** Assessment of TMEM240 antibody specificity (Santa Cruz Biotechnology) in anterior lobe sections.

(a1-a2) Positive control using only the TMEM240 antibody on sagittal cerebellar sections. Immunostaining of TMEM240 is shown in red. Competition studies were performed by incubating the TMEM240 antibody with the immunogenic peptide (b1-b2) or the human recombinant protein (c1-c2). A non-specific blocking peptide control experiment was performed by incubating the antibody with a non-competitive BDNF peptide (d1-d2). Images were obtained with a Zeiss Axio Imager Z2 microscope. Scale bar: 50 μm. ML: molecular layer, GL: granular layer, PL: Purkinje-cell layer

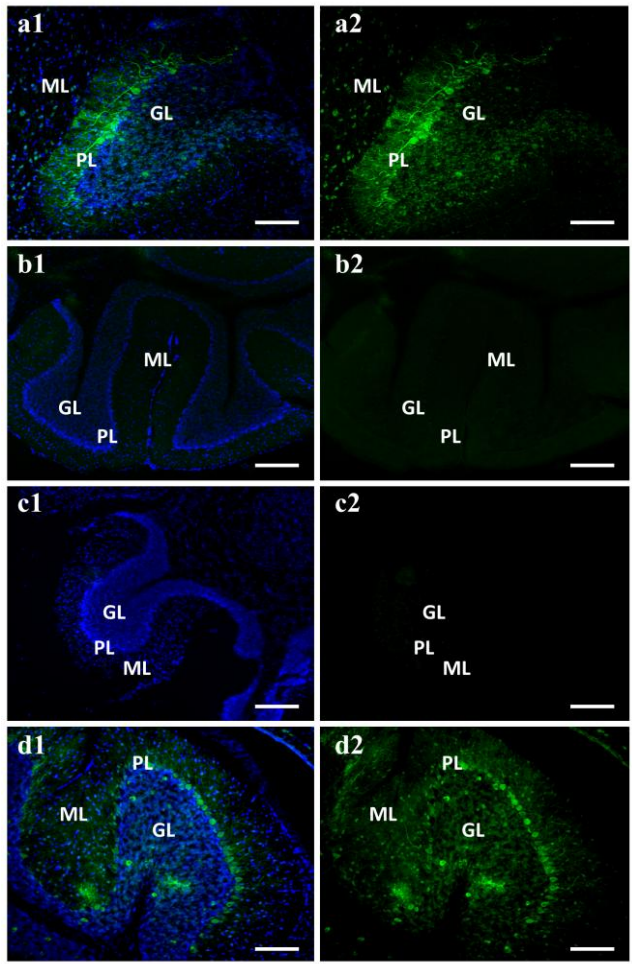

**Supplemental Figure S2** Assessment of TMEM240/63-77 antibody specificity) in anterior lobe sections.

(a1-a2) Positive control using only the TMEM240 antibody on sagittal cerebellar sections. Immunostaining of TMEM240 is shown in green. Competition studies were assessed by incubating the TMEM240 antibody with the immunogenic peptide (b1-b2) or the human recombinant protein (c1-c2). Non-specific blocking peptide control: incubation with a non-competitive BDNF peptide (d1-d2). Images were obtained with a Zeiss Axio Imager Z2 microscope. Scale bar: 50 μm. ML: molecular layer, GL: granular layer, PL: Purkinje-cell layer

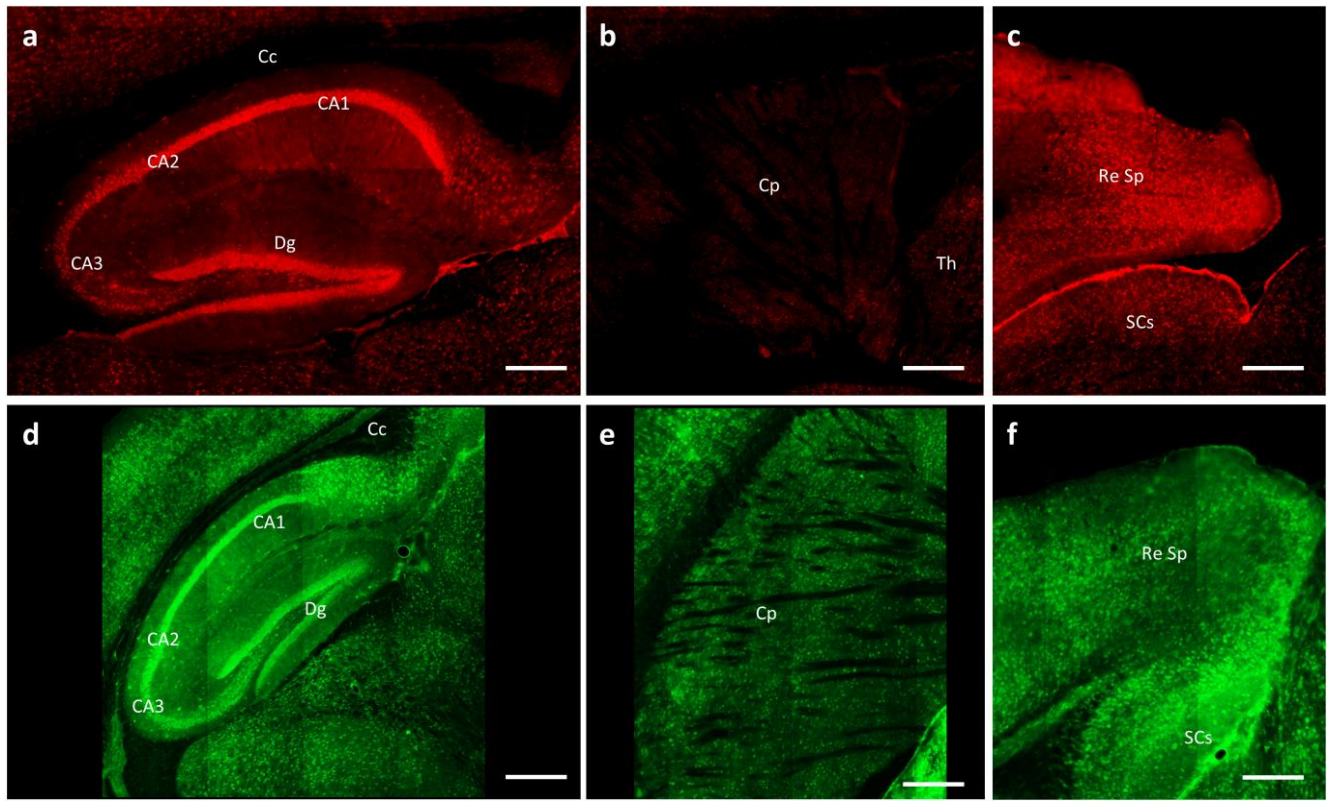

**Supplemental Figure S3** TMEM240 expression in the mouse brain.

TMEM240 immunoreactivity detected by immunofluorescence labeling (Santa Cruz Biotechnology, in red) in the hippocampal formation (a), caudoputamen (b), and retrospenial area of the cerebral cortex (c). Scale bars: 200 μm for the hippocampal formation and caudoputamen and 100 μm for the retrospenial area. TMEM240 immunoreactivity detected by immunofluorescence labeling (TMEM240/63-77, in green) in the hippocampal formation (d), caudoputamen (e), and retrospenial area of the cerebral cortex (f). Scale bars: 200 μm for the hippocampal formation and caudoputamen and 100 μm for the retrospenial area. Images were obtained by confocal microscopy., CA1, CA2, CA3: CA1, CA2, CA3 fields, Cc: corpus callosum, Cp: caudoputamen, Dg: dentate gyrus, Re Sp: retrosplenial area, SCs: superior colliculus sensory related area, TH: thalamus

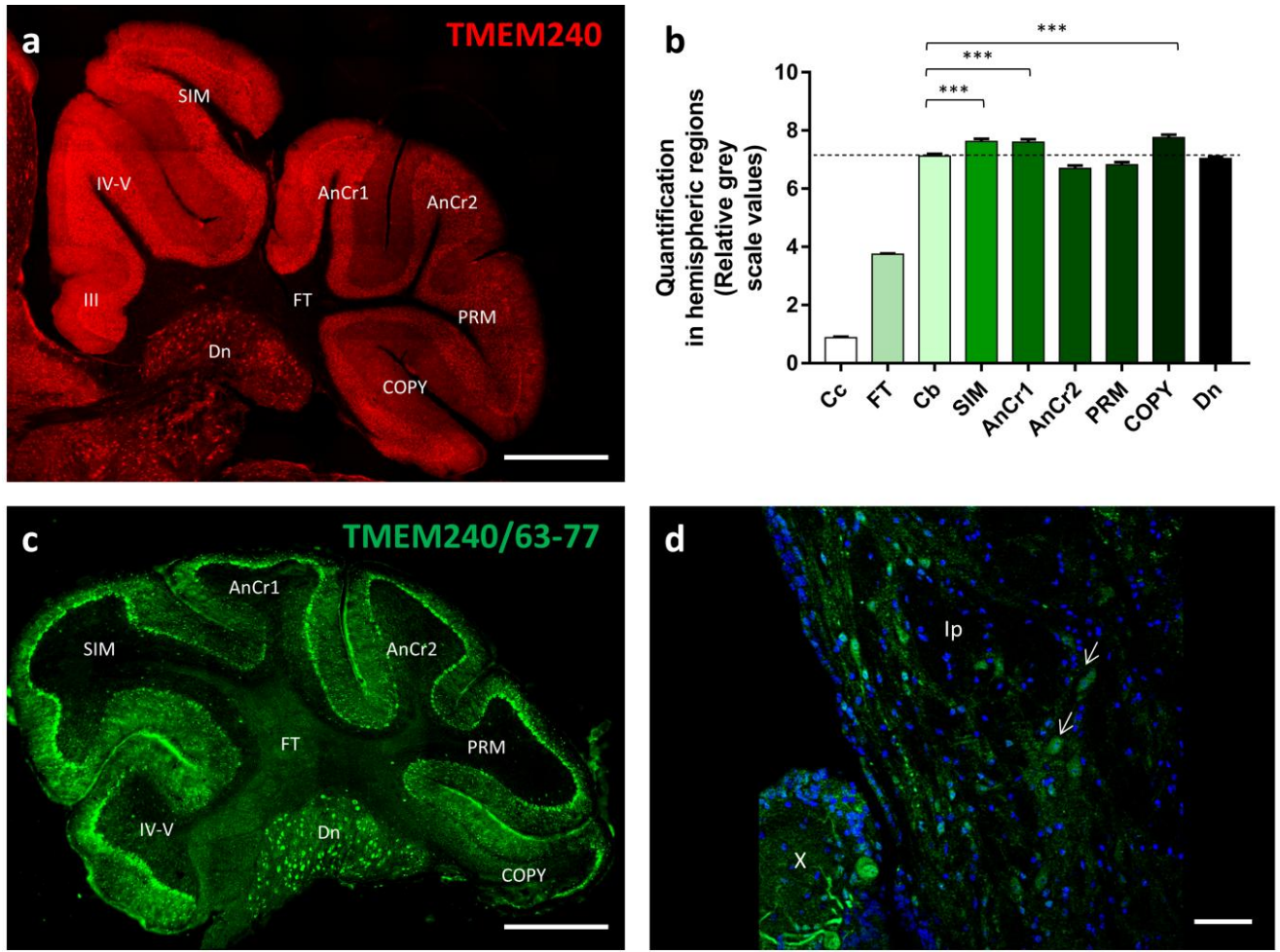

**Supplemental Figure S4** TMEM240 expression in cerebellar cortex and deep nuclei.

(a) TMEM240 immunoreactivity detected by immunofluorescence labeling (Santa Cruz Biotechnology, in red) of parasagittal cerebellar sections. Scale bar: 500 μm. (b) Immunofluorescence quantification of TMEM240 staining (Santa Cruz Biotechnology) of cerebellar lobules, cerebellar fiber tracts, and deep nuclei (expressed as the number of pixels, normalized to relative grey scale values. Total cerebellar staining (Cb) was used as the reference measurement for statistical analysis. $*p < 0.05$, $**p < 0.01$, $***p < 0.001$. (c) TMEM240 immunoreactivity detected by immunofluorescence labeling (TMEM240/63-77, in green) of parasagittal cerebellar sections. Scale bar: 500 μm. (d) Magnification of TMEM240 staining (TMEM240/63-77) in an interposed nucleus. Scale bar: 50 μm. White arrows depict neurons with a large cellular body. Images were obtained by confocal microscopy. AnCr1: ansiform Crus 1 lobule, AnCr2: ansiform Crus 2 lobule, Cb: cerebellum, Cc: corpus callosum, COPY: copula pyramidis, Dn: dentate nucleus, FT: cerebellar fiber tracts, Ip: interposed nucleus, PRM: paramedian lobule, SIM: simple lobule, III: central lobule III, IV-V: culmen (lobules IV-V), X: nodulus (lobule X)

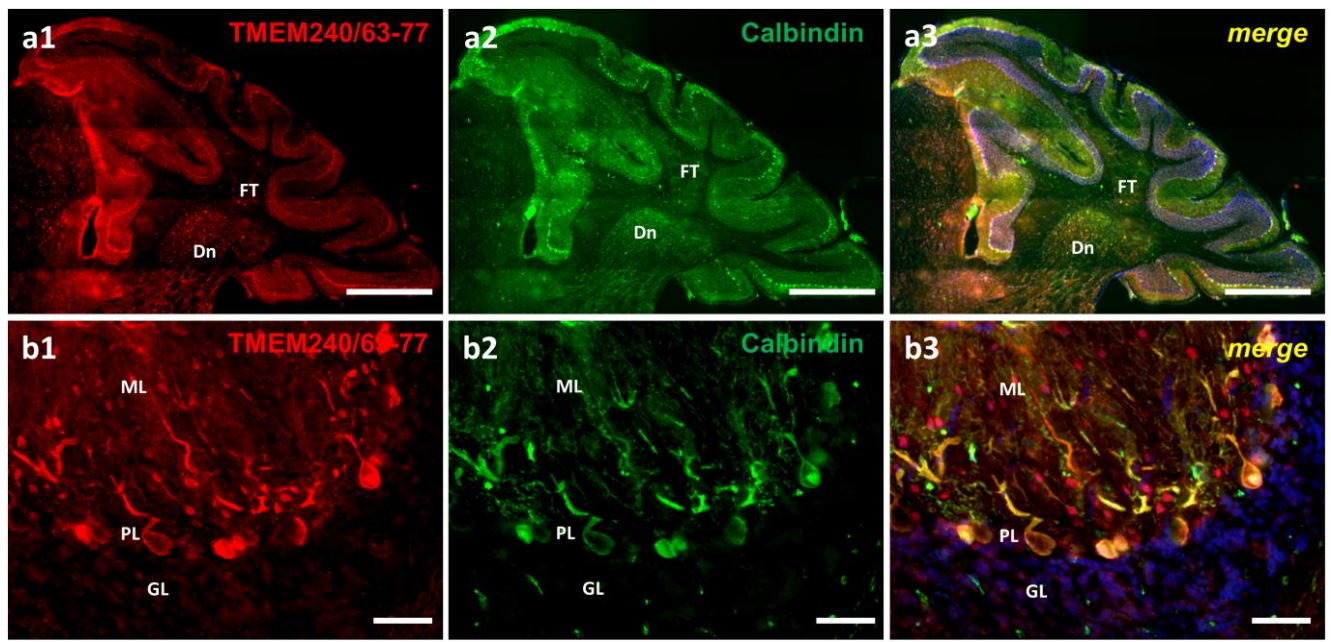

**Supplemental Figure S5** Expression of TMEM240 in Purkinje cells (TMEM240/63-77 antibody).

(a1-a3) Double immunofluorescence labeling of TMEM240 (TMEM240/63-77, in red) and calbindin (in green) in sagittal cerebellar sections. White arrows indicate TMEM240-positive cells in the white matter and deep nuclei. Scale bar: 500 μm. (b1-b3) Magnification of double immunofluorescence labeling image of TMEM240 (TMEM240/63-77, in red) and calbindin (in green) in sagittal cerebellar sections. TMEM240-positive cells corresponding to the labeling of basket cells (1) and stellate cells (2). Images were obtained by confocal microscopy. Scale bar: 50 μm. Dn: dentate nucleus, FT: cerebellar fiber tracts, ML: molecular layer, GL: granular layer, PL: Purkinje-cell layer

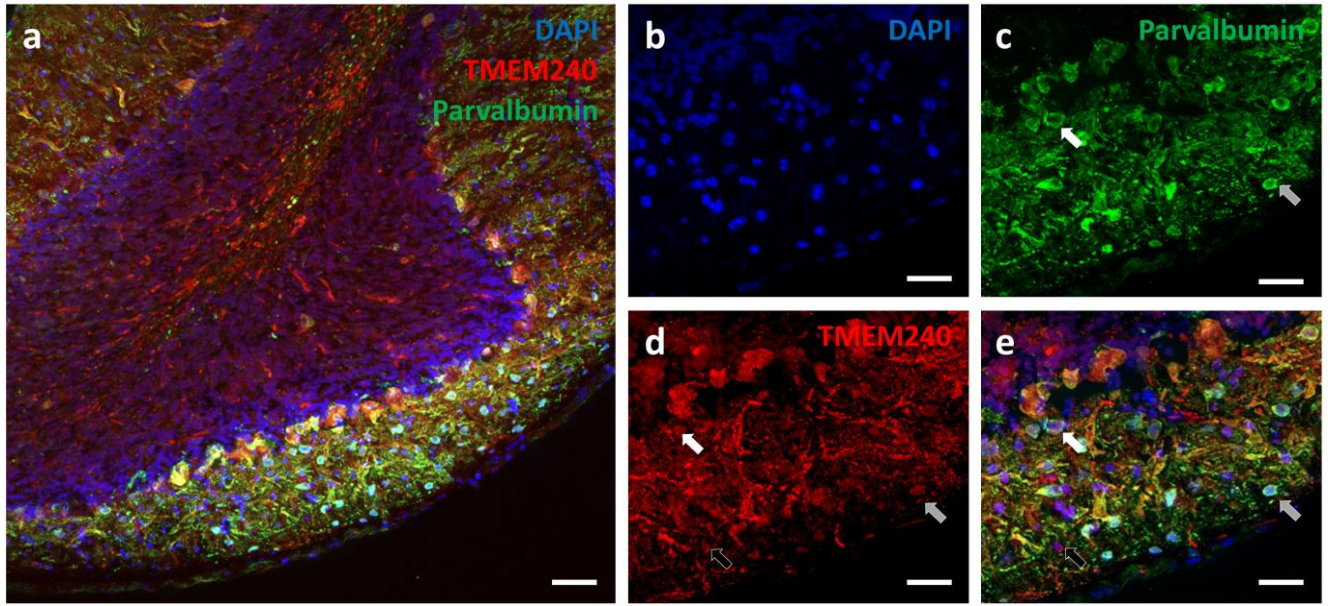

**Supplemental Figure S6** Expression of TMEM240 in interneurons.

(a) Double immunofluorescence labeling of TMEM240 (Santa Cruz Biotechnology, in red) and parvalbumin (in green) in sagittal cerebellar sections. Scale bar: 50 μm (b) Magnification of DNA labeling using DAPI. (c) Magnification of parvalbumin labeling (in green) in Purkinje and molecular layers. (d) Magnification of TMEM240 labeling (Santa Cruz Biotechnology, in red) in Purkinje and molecular layers. (e) Magnification of double labeling of TMEM240 and parvalbumin. The white arrow indicates basket cells, the grey arrow stellate cells, and the black arrow an excitatory neuron in the molecular layer. Scale bar: 20 μm. Images were obtained by confocal microscopy.

| Antibody | Host and clonality | Dilution | Secondary Alexa Fluor® | Saturation serum | Source and RRID number | Immunogen |
| --- | --- | --- | --- | --- | --- | --- |
| C1orf70 (G-16) | Goat Polyclonal | 1/200 | Donkey anti goat 568 nm | Donkey serum | sc-245675 (Santa Cruz Biotechnology) | Synthetic peptide corresponding to an internal region of human TMEM240 |
| TMEM240/63-77 | Rabbit Polyclonal | 1/1000 | Donkey anti rabbit 488 nm | Donkey serum | (Agro-Bio) | Synthetic peptide conjugated to KLH corresponding to amino acids 63 to 67 of human TMEM240 (PYDGDQSVVDASENY) |
| Calbindin | Mouse Monoclonal | 1/5000 | Goat anti mouse 488 nm | Normal Goat serum | (Swant) RRID: AB_10000347 | Purified Calbindin D-28k from chicken gut |
| Synaptophysin | Rabbit Polyclonal | 1/250 | Goat anti rabbit 488 nm | Normal Goat serum | ab14692 (Abcam) RRID: AB_301417 | Synthetic peptide corresponding to amino acids 41-62 of human Synaptophysin (FATCGSYSGELQLSVDCANKTE) |
| VGLUT1 | Guinea pig Polyclonal | 1/5000 | Goat anti guinea pig 488 nm | Normal Goat serum | AB5905 (Merck Millipore) RRID: AB_2301751 | Synthetic peptide c corresponding to amino acids 541-560 of rat VGLUT1 (YGATHSTVQPPRPPPPVRDY) |
| VGLUT2 | Guinea pig Polyclonal | 1/750 | Goat anti guinea pig 488 nm | Normal Goat serum | AB2251-I (Merck Millipore) RRID: AB_2665454 | Synthetic peptide conjugated to KLH corresponding to the C-terminal sequence of rat VGluT2 |
| Iba1 | Rabbit Polyclonal | 1/1000 | Donkey anti rabbit 488 nm | Donkey serum | (Wako) RRID: AB_839504 | Synthetic peptide corresponding to the Iba1 Carboxy-terminal sequence, which was conserved among human, rat and mouse Iba1 protein sequence |
| PSD95 (D74D3) | Rabbit Monoclonal | 1/200 | Goat anti rabbit 488 nm | Normal Goat serum | (Cell Signaling Technology) RRID: AB_1264242 | Synthetic peptide corresponding to the N-terminal sequence of human PSD95 |
| GFAP (2E1) | Mouse Monoclonal | 1/1000 | Donkey anti mouse 488 nm | Donkey serum | sc-33673 (Santa Cruz Biotechnology) RRID: AB_627673 | Homogenate of bovine spinal chord |
| Parvalbumin | Rabbit polyclonal | 1/500 | Donkey anti rabbit 488 nm | Donkey serum | ab256811 (Abcam) | Recombinant fragment within the center region of mouse parvalbumin |

**Supplemental Table S1** Antibodies used in immunohistochemical experiments

| Antibody | Host and clonality | Dilution | Molecular Weight (kDa) | Saturation serum | Source and RRID number | Immunogen |
|---|---|---|---|---|---|---|
| C1orf70 (G-16) | Goat Polyclonal | 1/1000 | 19.9 | BSA 5% | sc-245675 (Santa Cruz Biotechnology) | Synthetic peptide corresponding to an internal region of human TMEM240 |
| TMEM240/63-77 | Rabbit Polyclonal | 1/1000 | 19.9 | BSA 5% | (Agro-Bio) | Synthetic peptide conjugated to KLH corresponding to amino acids 63 to 67 of human TMEM240 (PYDGDQSVVDASENY) |
| PSD95 (D74D3) | Rabbit Monoclonal | 1/1000 | 95 | Milk 5% | (Cell Signaling) #3409 | Synthetic peptide corresponding to the amino-terminal sequence of human PSD95 |
| β-Actin Clone AC-15 | Mouse Monoclonal | 1/10000 | 42 | Milk 5% | #A5441 (Sigma-Aldrich) | Synthetic β-cytoplasmic actin N-terminal peptide |
| SNAP25 (H-1) | Mouse Monoclonal | 1/1000 | 28 | Milk 5% | sc-376713 (Santa Cruz Biotechnology) | Synthetic peptide corresponding to amino acids raised against amino acids 91-140 of human SNAP25 |

**Supplemental Table S2** Antibodies used in western-blotting experiments

| Brain regions | Brain areas | Areas subparts | Expression |
|---|---|---|---|
| **Cerebral nuclei** | Striatum | Caudoputamen | + |
| | | Nucleus accumbens | ++ |
| | | Olfactory tubercle | + |
| | Pallidum | Bed nuclei of the stria terminalis | + |
| | | Substantia innominata | + |
| | | Diagonal band nucleus | + |
| **Olfactory areas** | Main olfactory bulb | | ++ |
| | Accessory olfactory bulb | | + |
| | Anterior olfactory nucleus | | + |
| | Piriform areas | | ++ |
| | Olfactory tubercle | | ++ |
| **Isocortex** | Orbital areas | | ++ |
| | Somatomotor related areas | | +++ |
| | Somatosensory related areas | | +++ |
| | Posterior parietal association areas | | +++ |
| | Visual areas | | +++ |
| | Retrosplenial area | | +++ |
| **Midbrain** | | Inferior colliculus | ++ |
| | | Superior colliculus sensory related | +++ |
| | | Superior colliculus motor related | ++ |
| | | Periaqueductal gray | + |
| | | Midbrain reticular nucleus | + |
| | | Red nucleus | ++ |
| | | Ventral tegmental area | +++ |
| **Interbrain** | Thalamus | Lateral habenula | + |
| | | Mediodorsal nucleus of thalamus | + |
| | | Ventral medial nucleus of thalamus | + |
| | | Parafascicular nucleus | + |
| | | Antero medial nucleus | + |
| | | Anterodorsal nucleus | + |
| | Hypothalamus | Medial preoptic area | + |
| | | Medial preoptic nucleus | + |
| | | Zona incerta | + |
| | | Paraventricular hypothalamic nucleus | + |
| | | Anterior hypothalamic nucleus | + |
| | | Dorsomedial nucleus of the hypothalamus | + |
| | | Ventromedial hypothalamic nucleus | +++ |
| | | Retrochiasmatic area | +++ |
| | | Posterior hypothalamic nucleus | ++ |
| | | Ventral premammillary nucleus | + |
| | | Dorsal premammillary nucleus | + |
| | | Medial mammillary nucleus | +++ |
| | | Supramammillary nucleus | +++ |
| **Hippocampal formation** | CA1 | Stratum lacunosum moleculare | ++ |
| | | Stratum radiatum | ++ |
| | | Pyramidal layer | +++ |
| | | Stratum oriens | ++ |
| | CA2 | Stratum lacunosum moleculare | ++ |
| | | Stratum radiatum | ++ |
| | | Pyramidal layer | +++ |
| | | Stratum oriens | ++ |
| | CA3 | Stratum lacunosum moleculare | + |
| | | Stratum radiatum | + |
| | | Pyramidal layer | + |
| | | Stratum oriens | + |
| | Subiculum | | ++ |
| | Dentate gyrus | Polymorph layer | ++ |
| | | Granule cell layer | +++ |
| | | Molecular layer | ++ |
| **Cerebellum** | Cerebellar cortex | Lingula I | +++ |
| | | Central lobule II | +++ |
| | | Central lobule III | +++ |
| | | Culmen IV-V | +++ |
| | | Declive VI | +++ |
| | | Folium tuber vermis VII | +++ |
| | | Pyramus VIII | +++ |
| | | Uvula IX | +++ |
| | | Nodulus X | +++ |
| | Deep nuclei | Dentate nucleus | +++ |
| | | Interposed nucleus | ++ |
| | | Fastigial nucleus | ++ |
| **Hindbrain** | Pons | Pontine gray | ++ |
| | | Superior olivary complex | + |

|  |  |  | Pontine reticular nucleus | + |
|---|---|---|---|---|
|  |  |  | Pontine reticular nucleus caudal part | + |
|  |  |  | Tegmental reticular nucleus | ++ |
|  |  |  | Laterodorsal tegmental nucleus | + |
|  |  | Medulla | Medial vestibular nucleus | + |
|  |  |  | Spinal vestibular nucleus | ++ |
|  |  |  | Nucleus of the solitary tract | + |
|  |  |  | Medullary reticular nucleus | + |
|  |  |  | Gigantocellular reticular nucleus | + |
|  |  |  | Intermediate reticular nucleus | + |
| **Fiber tracts** | Corpus callosum |  |  | - |
|  | Cerebellar fiber tracts |  |  | + |
|  | Solitary tract |  |  | - |
|  | Anterior commissure |  |  | - |
|  | Fornix system |  |  | - |
|  | Stria medullaris |  |  | - |
|  | Cerebral peduncle |  |  | - |

**Supplemental Table S3** Summary of TMEM240 expression in the adult mouse brain regions, areas and subparts.